\documentclass{article}

\usepackage{authblk}
\usepackage{amsmath,amssymb,amsthm}
\usepackage{graphicx}
\usepackage{setspace}
\usepackage{lineno}
\usepackage[margin=1in]{geometry}
\usepackage{float}
\usepackage{hyperref}
\usepackage{comment}

\usepackage{enumerate}

\usepackage{tikz}
\usetikzlibrary{arrows.meta,positioning,calc,decorations.pathmorphing}

\large

\title{
Symmetric Locality as a Tetrahedron: \\
A Symmetry-Reduced Geometric Representation \\
of the $(3,3,2,2)$ Bell Scenario
}

\author[1]{Marek Gazdzicki}
\author[1]{Francesco Giacosa}
\author[2]{Pawel Piesowicz}
\affil[1]{Jan Kochanowski University, Kielce, Poland}
\affil[2]{Speightstown, Barbados}

\begin{document}
\maketitle

\begin{abstract}
We present a geometric characterisation of local models in the bipartite Bell scenario with three measurement settings per site
and binary outcomes, i.e.\ the $(3,3,2,2)$ case.
Restricting attention to indistinguishable sites, thus resulting in a `symmetric local' scenario, we introduce a three-dimensional mixed-moment space in which the mixed moments are calculated under off-diagonal measurement settings.

In this reduced representation, the symmetric local region assumes the remarkably simple form of a regular tetrahedron - the 'pyramid'.
We prove that only three inequalities are required to characterise the interior of this
region. We call them the pyramid inequalities that separate symmetric local ($\mathcal{SL}$) models from their complement, non-symmetric local ($\mathcal{\overline{SL}}$) models. Remarkably, in this representation, the 'symmetric quantum' region forms an elliptope defined by a single inequality.
We also clarify the relation between the symmetry-reduced pyramid
representation and the full $(3,3,2,2)$ Bell polytope in the
36-dimensional conditional-probability space, which possesses
684 facet-defining inequalities.
The reduction from 684 to three reflects normalisation, symmetry
reduction, and projection to the mixed-moment space.

In the pyramid representation, the hierarchy
$\mathcal{SL} \subsetneq \mathcal{SQ} \subsetneq \mathcal{NS}$ appears geometrically as a tetrahedron embedded in a somewhat larger
curved body of symmetric quantum models, $\mathcal{SQ}$, which in turn is embedded in a cube of no-signalling models, $\mathcal{NS}$. This provides a unified three-dimensional visualisation of symmetric local, symmetric quantum, and symmetric post-quantum correlations.
The qualitative advantages of the pyramid representation over the standard one-dimensional CSHS representation of the $(2,2,2,2)$ case are also discussed.

\end{abstract}

\section{Introduction}
\label{sec:introduction}

The origin of Bell-type limits (inequalities) on correlations in local models can be traced back to the Einstein--Podolsky--Rosen (EPR) paradox. 
EPR argued that a complete physical theory should admit a
local-realistic description, where outcomes are determined by
pre-existing properties, some of which may be hidden from us. 
EPR refused to accept instantaneous influences propagating
between spatially separated systems as suggested by quantum mechanics. 
Einstein famously called them
``spukhafte Fernwirkung'' (``ghostly action at a distance'').

Bell’s theorem demonstrates that correlations between remote events predicted by quantum mechanics
cannot, in general, be reproduced by local models defined by
Bell factorisation~\cite{Bell:1964kc,Bell:1980wg}, see Refs.~\cite{Gisin:2007gps,Horodecki:2009zz} for a review and Refs.~\cite{Silagadze:2009mi,Alford:2015xpa} for a pedagogical introduction.
Since the pioneering experiments of Aspect and collaborators in the
1980s~\cite{Aspect:1982}, and culminating in tests
performed in 2015~\cite{Hensen:2015,Giustina:2015,Shalm:2015,Brunner:2013est,Larsson:2014rzp}, violations of Bell inequalities have been firmly confirmed experimentally.
These results rule out local hidden-variable descriptions of nature and
provide the basis for device-independent quantum information
protocols~\cite{Brunner:2013est}. 
Alternative approaches that relax some of the assumptions underlying
Bell’s theorem have also been discussed in the literature.
In particular, superdeterministic models abandon the
assumption of independence of measurement settings
and the hidden variables~\cite{Bell:1977,tHooft:2007oio,Hall:2010zzf,Hossenfelder:2019shy}.
While Bell factorisation may still hold in such models, they lie outside the standard framework of local hidden-variable theories considered here.
Also, note that the nature of measurement within quantum mechanics, whether described by purely unitary evolution or involving an actual collapse of the wave function, remains unsettled; for concise overviews, see e.g. Refs.~\cite{Bassi:2003gd,Giacosa:2014kay}.

In the bipartite scenario with $m$ measurement settings per site and binary
outcomes ($n=2$), the set of local models forms a convex
polytope in the space of conditional probabilities.
Its facets correspond to Bell inequalities. 
For the simplest $(2,2,2,2)$ scenario, the local polytope is 
characterised (up to trivial changes of notation and conventions) by the CHSH inequality~\cite{CHSH1969}.
In general, the geometry of Bell polytopes and their facet structure
has been extensively studied; see, for example,
Refs.~\cite{Brunner:2013est,Froissart:1981ni}.

The Bell inequalities constrain the correlations allowed by local models. 
Similarly, as shown by Tsirelson~\cite{Tsirelson:1980}, the correlations due to quantum models are also limited.  
The corresponding bounds follow from the fact that 
observables admitting a quantum realisation
can be represented as scalar products of real unit vectors~\cite{Tsirelson:1980}.

When the number of settings per site increases to three,
the $(3,3,2,2)$ scenario exhibits a much richer structure.
In the full conditional-probability space, the local polytope possesses 684 facet-defining inequalities.
Up to relabelings and symmetry transformations, these facets
belong to two inequivalent families of non-trivial Bell
inequalities: the CHSH-type inequalities and the
$I_{3322}$ (Froissart-type) inequalities~\cite{Brunner:2013est,Froissart:1981ni}.

In the present work, we restrict attention to indistinguishable sites and introduce a
symmetry-reduced mixed-moment space $(X,Y,Z)$ constructed
under conditions of off-diagonal measurement settings. We thus call this system 'symmetric local'. 
Then we show that the
$(3,3,2,2)$ symmetric local region becomes a regular tetrahedron,
with its interior fully characterised by three inequalities.
The pyramid representation provides a simple and geometric test of
locality, complementary to the standard full-polytope analysis.
Moreover, in the mixed-moment space, the set of symmetric quantum models
appears as the elliptope, containing the symmetric local tetrahedron and
embedded in the cube of no-signalling models.

Symmetry-based reductions of Bell scenarios have previously been explored in the literature. Bancal \emph{et al.}~\cite{Bancal:2010vhq} introduced a general framework for projecting Bell polytopes onto subspaces invariant under permutations of the parties and showed how symmetric Bell inequalities can be obtained from the resulting reduced polytopes. Their approach was subsequently extended in several directions, including the study of multipartite Bell inequalities and more general symmetry groups~\cite{Tura:2013qod,Tura:2013qod}. Another example is provided by Cai \emph{et al.}~\cite{Cai:2016ech}, who employed a depolarisation procedure that preserves the value of the CGLMP expression while projecting the correlation space onto a three-dimensional subspace. More recently, symmetry methods have also been used to simplify the computation of quantum bounds and semidefinite relaxations of Bell scenarios~\cite{Tavakoli:2021vcg}. These works demonstrate that symmetry can substantially reduce the complexity of Bell-type probability spaces and reveal their geometric structure.
After accounting for the opposite outcome-sign convention, the same tetrahedron–elliptope–cube geometry was obtained by Janas~\emph{et al.}~\cite{Janas:2021xyg}  for Mermin’s three-setting scenario~\cite{Mermin1981-MERQMF} under perfect same-setting anticorrelation.

The present work differs from these approaches in an important respect. Previous symmetry reductions were imposed at the level of the \emph{observed} conditional probabilities, i.e.\ on probability distributions after the hidden-variable decomposition has been averaged over the hidden-variable states. In contrast, we impose exchange symmetry directly at the level of the hidden-variable response functions,
which is a stronger requirement than exchange symmetry of the observed probabilities alone. The hidden-variable symmetry implies the corresponding symmetry of the observable conditional probabilities, but the converse is generally not true. Combined with Bell factorisation and the projection from the conditional-probability space to the space of off-diagonal mixed moments, this stronger symmetry requirement leads to the tetrahedral geometry of the symmetric local region derived in this work.

The paper is organised as follows.
Section~\ref{sec:setup} presents the setup and the definition of symmetric local models.
The symmetry-reduced mixed moment space is introduced in Sec.~\ref{sec:tetrahedron}, where it is also shown that in this space the set of symmetric local models is a regular tetrahedron.
In Sec.~\ref{sec:number}, the number of Bell's inequalities is counted in the mixed moment space and related to the corresponding number in the full conditional-probability space.
The physics discussion presented in Sec .~\ref{sec:physics} includes the presentation of the set of quantum and non-signalling models in the mixed-moment space.
The summary given in Sec.~\ref{sec:summary} closes the paper.
The main text is complemented by appendices.

\section{The setup and symmetric locality}
\label{sec:setup}

\begin{figure}[H]
\centering
\includegraphics[width=0.8\linewidth]{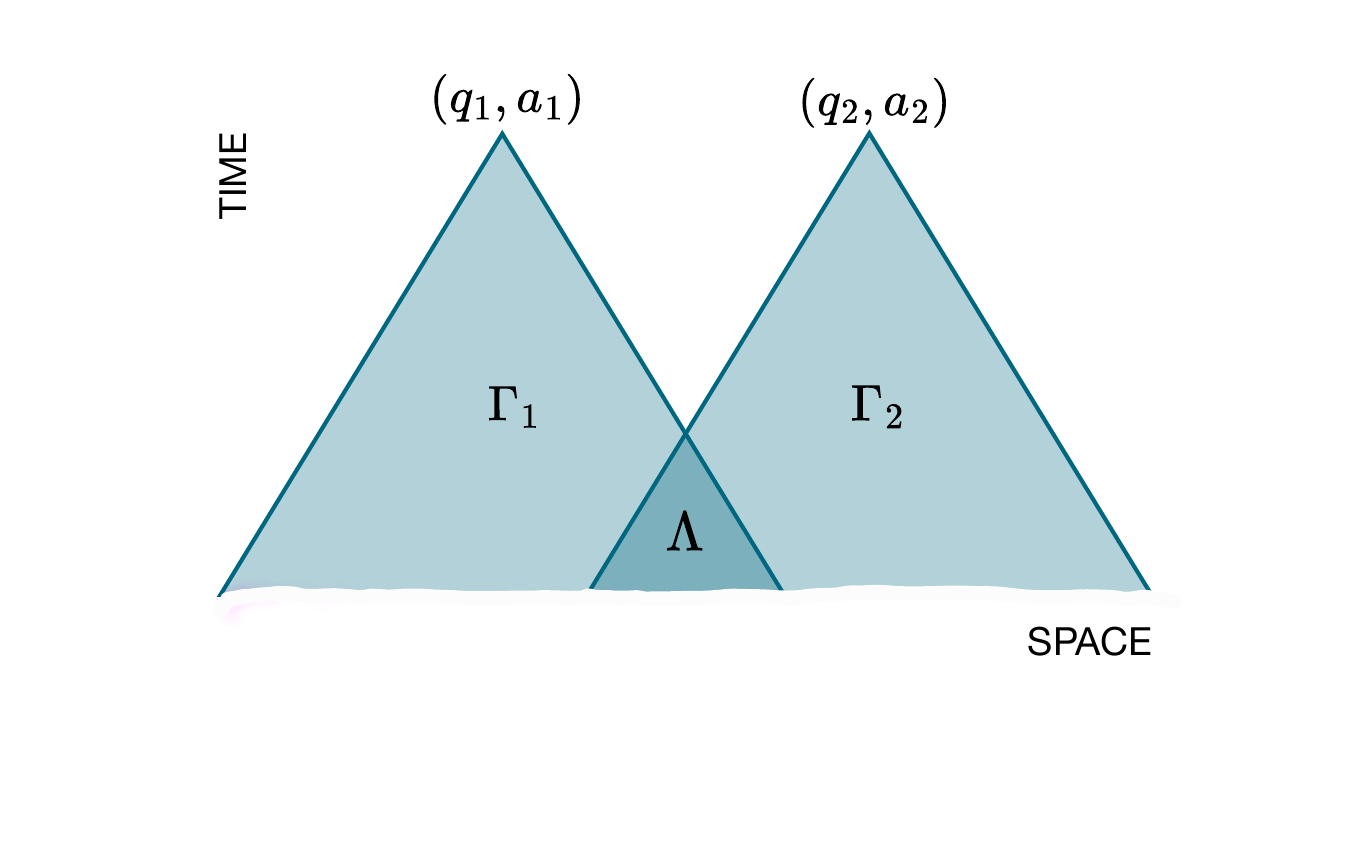}
\vspace*{-1cm}
\caption{Light-cone diagram for two sites, $(q_1, a_1)$ and $(q_2, a_2)$ and their corresponding past domains, $\Gamma_1$ and $\Gamma_2$. The common past events located in the space-time region $\Lambda$ may have correlated
$a_1$ and $a_2$. Time flows upwards. See text for further details.}
\label{fig:lightcone}
\end{figure}

We consider two spatially separated sites (labelled 1 and 2) at time $t$.
Each site is equipped with a device that measures the properties of objects there.
Within the setup $q_1, q_2$ denote the settings labels (``questions'')
at the two sites. We assume three possible settings,  $q_1, q_2 \in \{0, 1, 2\}$,
whereas $a_1, a_2 \in \{+1,-1\}$ denote the corresponding labels of binary outcomes (``answers'').
The setup is schematically presented in Fig.~\ref{fig:lightcone}.

A probabilistic model is specified by the conditional joint distribution
\begin{equation}
\label{eq:joint}
P(a_1,a_2 \mid q_1,q_2)~,
\end{equation}
which is assumed to be independent of time.

Let $\lambda$ denote a hidden variable whose value is set in the common past $\Lambda$ of the light cones 
of the two sites, $\Gamma_1$ and $\Gamma_2$.
We assume measurement independence (sometimes called the `free-will'
assumption~\cite{tHooft:2007oio}), namely that the measurement settings are statistically
independent of the hidden variable established in the common past:
\begin{equation}
h(q_1,q_2 \mid \lambda)=h(q_1,q_2)~,
\label{eq:measurement_independence}
\end{equation}
where $h(...)$ is the probability distribution of $q_1, q_2$.
Using Bayes' theorem, this condition is equivalent to
$f(\lambda \mid q_1,q_2)=f(\lambda)$ for non-vanishing probabilities
$h(q_1,q_2)$\footnote{Although measurement independence does not formally require the settings
to be random, independently generating the settings event-by-event makes
their statistical independence from the hidden variable substantially
more plausible experimentally. In contrast, one may expect that a fully predetermined setting sequence is more likely correlated with the hidden
variables through common past causes. Whatever the procedure is, the superdeterministic loophole remains logically open in all Bell experiments, because the statistical independence of the measurement
settings from the hidden variables cannot be established with absolute
certainty.}.
This free-will assumption
excludes superdeterministic models~\cite{Bell:1977,tHooft:2007oio,Hall:2010zzf,Hossenfelder:2019shy}.
For definiteness, in the following, we consider discrete values $\lambda=1,...,N_{\lambda}$, leading to the probability distribution normalisation
\[
\sum_{\lambda=1}^{N_\lambda}f(\lambda)=f_1+...f_{N_{\lambda}}=1~,
\]
with $f(\lambda)=f_{\lambda}$. The extension to (un)countable infinite $\lambda$ is straightforward.  Both the hidden variable \(\lambda\) and its distribution \(f(\lambda)\) are
assumed to be experimentally inaccessible.

In general, local models are then defined by Bell factorisation
~\cite{Bell:1964kc,Bell:1980wg} as,
\begin{equation}
p(a_1,a_2 \mid q_1,q_2,\lambda)
=
p_1(a_1 \mid q_1,\lambda)\,
p_2(a_2 \mid q_2,\lambda)~.
\label{eq:bell_factorisation}
\end{equation}
Note that here locality is understood in the sense of Bell locality (or local causality), which implies the factorisation of probabilities above. We stress, however, that the term locality has a different meaning in quantum field theory, where it refers to local interactions and, more generally, to microcausality, namely the commutativity of space-like-separated observables. Relativistic QFT satisfies locality in this sense, yet it violates Bell locality through the breakdown of probability factorisation for entangled states. To avoid this ambiguity, some authors, e.g. \cite{Alford:2015xpa}, refer to Bell locality as strong locality.

Averaging over the hidden variable in Eq. \ref{eq:bell_factorisation} gives the observed distribution for the local models:
\begin{equation}
P(a_1,a_2 \mid q_1,q_2)
=
\sum_{\lambda}
p_1(a_1 \mid q_1,\lambda)\,
p_2(a_2 \mid q_2,\lambda)\,
f(\lambda)~.
\label{eq:SL_average}
\end{equation}

Importantly, in what follows, we consider only the indistinguishable sites 1 and 2. 
Thus, the setting and outcome label distributions are the same at both sites for every value of the hidden variable:
\begin{equation}
\label{eq:symm_b}
p_1(a \mid q,\lambda)=p_2(a \mid q,\lambda)\equiv p(a \mid q,\lambda)~.
\end{equation}
Naturally, one assumes here that the correspondence between the numerical labels $\{0,1,2\}$, $\{-1,1\}$ and the physical settings and outcomes, respectively, is identical at both sites. This condition is the precise definition of a \textit{symmetric local} setup.
From the symmetry condition of Eq. ~\eqref{eq:symm_b} follows the general symmetry of the observed distribution:
\begin{equation}
\label{eq:symm_g}
P(a_1,a_2\mid q_1,q_2)
=
P(a_2,a_1\mid q_2,q_1)~.
\end{equation}
for any values of 
$a_1, a_2$ and $q_1, q_2$.
For indistinguishable sites, the general symmetry of the observed distribution~\eqref{eq:symm_g} has to be fulfilled by any model, independently of whether it obeys Bell's factorisation.

\section{The symmetric local $\mathcal{SL}$ models as regular tetrahedron}
\label{sec:tetrahedron}
\subsection{Definition of mixed moments}
\label{sec:N0}
Let us define an event quantity as a product of outcomes at sites 1 and 2 as:
\begin{equation}
a_1 \cdot a_2~,
\end{equation}
and then, off-diagonal mixed moments by averaging the event quantity over events with off-diagonal values of $q_1, q_2$:
\begin{align}
\label{eq:six}
M_{01} = \langle a_1 a_2 \rangle_{q_1=0,\,q_2=1}~,\quad  &
M_{10} = \langle a_1 a_2 \rangle_{q_1=1,\,q_2=0}~,\quad  \nonumber \\
M_{02} = \langle a_1 a_2 \rangle_{q_1=0,\,q_2=2}~,\quad  &
M_{20} = \langle a_1 a_2 \rangle_{q_1=2,\,q_2=0}~,\quad  \nonumber \\
M_{12} = \langle a_1 a_2 \rangle_{q_1=1,\,q_2=2}~,\quad  &
M_{21} = \langle a_1 a_2 \rangle_{q_1=2,\,q_2=1}~.\quad
\end{align}

The assumed symmetry of the observed distribution in Eq. ~\eqref{eq:symm_g} implies:
\begin{equation}
\begin{aligned}
\langle a_1 a_2 \rangle_{q_1\,q_2}
&= \sum_{a_1}\sum_{a_2} a_1 a_2 \, P(a_1,a_2 \mid q_1, q_2) \\
&= \sum_{a_1}\sum_{a_2} a_1 a_2 \, P(a_2,a_1 \mid q_2, q_1)
= \langle a_1 a_2 \rangle_{q_2\,q_1}~.
\end{aligned}
\label{eq:example}
\end{equation}
This reduces the description from six moments in Eq. ~\eqref{eq:six} to three:
\begin{align}
X=M_{01}=M_{10}~,\quad  \nonumber \\
Y=M_{02}=M_{20}~,\quad  \nonumber \\
Z=M_{12}=M_{21}~.\quad
\label{eq:mm}
\end{align}

In the following, we determine the region in $(X,Y,Z)$ space accessible to $\mathcal{SL}$ models.

\subsection{The case $N_\lambda=1$: curved-side pyramid}
\label{sec:N1}

For a single value of the hidden variable ($N_{\lambda}=1$ and $f_1=1$),  averaging over $\lambda$ in Eq. \eqref{eq:SL_average} is trivial. Thus, the joint probability distribution factorises~\eqref{eq:joint} directly,
\begin{equation}
P(a_1,a_2\mid q_1,q_2)
=
P(a_1\mid q_1)\,
P(a_2\mid q_2)~,
\label{eq:n1_fact}
\end{equation}
then
\begin{equation}
\langle a_1 a_2 \rangle_{q_1\,q_2}
=
\langle a_1\rangle_{q_1}\,
\langle a_2\rangle_{q_2}~.
\end{equation}
Given indistinguishable sites 1 and 2 defined by Eq.~\eqref{eq:symm_b}, we have $\langle a_1 \rangle_{q_1=0} = \langle a_2 \rangle_{q_2=0}$ and so on. So we can simplify the notation by reducing the subscripts and defining:
\begin{equation}
\alpha=\langle a\rangle_{0}~,\quad
\beta=\langle a\rangle_{1}~,\quad
\gamma=\langle a\rangle_{2}~,
\end{equation}
with $-1\le\alpha,\beta,\gamma\le1$,
one gets
\begin{align}
X &= \alpha \cdot \beta~, \nonumber \\
Y &= \alpha \cdot \gamma~, \nonumber \\
Z &= \beta \cdot \gamma~,
\label{eq:xyz_param}
\end{align}

\begin{figure}[!t]
\centering
\includegraphics[width=0.45\linewidth]{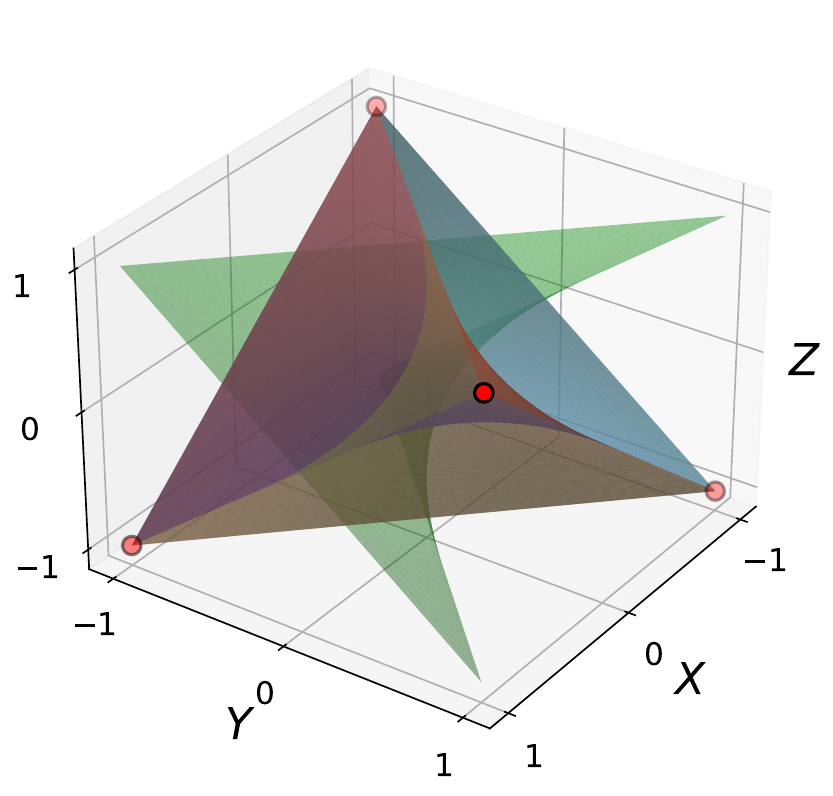}
\includegraphics[width=0.45\linewidth]{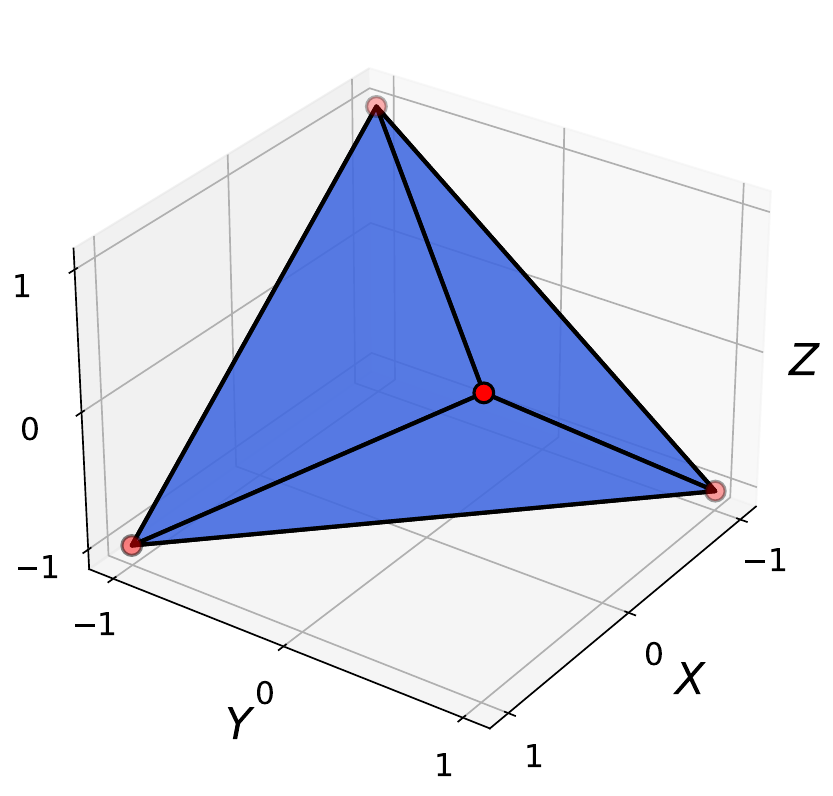}
\caption{
Geometry of the symmetric local models in the symmetry-reduced
mixed-moment space $(X,Y,Z)$.
\textit{Left:}
The curved closed surface forms the boundary of the $N_\lambda=1$ factorisable
models parametrised by
$X=\alpha\beta$, $Y=\alpha\gamma$, $Z=\beta\gamma$ with
$|\alpha|,|\beta|,|\gamma|\le1$.
It consists of the six hyperbolic faces arising from the boundaries
$|\alpha|=1$, $|\beta|=1$, and $|\gamma|=1$ (each shown in a different colour).
The four red vertices, $(1,1,1)$, $(-1,-1,1)$, $(-1,1,-1)$, and $(1,-1,-1)$, mark the extremal deterministic models.
\textit{Right:}
Convex mixtures of such models generate the full set of
$\mathcal{SL}$ models forming the regular
tetrahedron with vertices.
The planar faces of the tetrahedron correspond to the three
inequalities defining the $\mathcal{SL}$ region.
}
\label{fig:tetrahedron}
\end{figure}

It is not difficult to see that
the image of Eq.~\eqref{eq:xyz_param} forms a tetrahedron-like structure.
The extreme points of the structure correspond to the deterministic models ($\alpha, \beta, \gamma = \pm1$).
They constitute four vertices of the structure:
\begin{align}
V_1 &= (1,1,1)~, \nonumber \\
V_2 &= (-1,-1,1)~, \nonumber \\
V_3 &= (-1,1,-1)~, \nonumber \\
V_4 &= (1,-1,-1)~.
\label{eq:vertices}
\end{align}
They are affinely independent and pairwise equidistant.

A useful geometric identity follows directly from the parametrisation of Eq.~\eqref{eq:xyz_param}:
\begin{equation}
XYZ=(\alpha\beta\gamma)^2~.
\end{equation}
Since $|\alpha|,|\beta|,|\gamma|\le1$, the right-hand side is non-negative and bounded by unity. Thus, the image of the mapping is constrained by a quadratic relation that bends the boundary surfaces inward relative to the planar tetrahedral facets. Geometrically, this explains why the $N_\lambda=1$ region forms a tetrahedron-like body with hyperbolically concave faces, as illustrated in 
Fig.~\ref{fig:tetrahedron}.

\subsection{General symmetric local models: convexity}
\label{sec:Noo}

As follows from the previous subsection, for each fixed value of $\lambda$ the corresponding point lies within the
tetrahedron-like region generated by Eq.~\eqref{eq:xyz_param}. The averaging over $\lambda$ in
Eq.~\eqref{eq:SL_average} with probability distribution $f(\lambda)$ and under the symmetry condition~\eqref{eq:symm_b}, corresponds to forming convex combinations
of such points. Consequently, the full set of symmetric local models is given
by the convex hull of the four vertices $V_1, V_2, V_3, V_4$ defined in Eq.~\eqref{eq:vertices}.
Thus, in the symmetry-reduced $(X,Y,Z)$ space, the $\mathcal{SL}$ region is a regular
tetrahedron with planar faces.

As an illustrative case, we consider $N_\lambda=2$. Using the probabilities $f_{1,2}$ with $f_1 + f_2 =1$, we obtain
\begin{equation}
    X=f_1X_1+f_2X_2~,\qquad 
    Y=f_1Y_1+f_2Y_2~,\qquad 
    Z=f_1Z_1+f_2Z_2~,
\end{equation}
where
\begin{align}
X_1 &= \alpha_1 \beta_1~, \nonumber\\
Y_1 &= \alpha_1 \gamma_1~, \nonumber\\
Z_1 &= \beta_1 \gamma_1~,
\end{align}
and similarly for $(X_2,Y_2,Z_2)$. Here,
\begin{align}
\alpha_1 &= \langle a_1 \rangle_{q_1=0,\lambda=1}
=
\langle a_2 \rangle_{q_2=0,\lambda=1}~, \nonumber\\
\beta_1 &= \langle a_1 \rangle_{q_1=1,\lambda=1}
=
\langle a_2 \rangle_{q_2=1,\lambda=1}~, \nonumber\\
\gamma_1 &= \langle a_1 \rangle_{q_1=2,\lambda=1}
=
\langle a_2 \rangle_{q_2=2,\lambda=1}~.
\end{align}
Similar expressions define $\alpha_2, \beta_2, \gamma_2$  for $\lambda=2$.

Since each fixed-$\lambda$ point $(X_\lambda,Y_\lambda,Z_\lambda)$ is of the form
\begin{equation}
(X_\lambda,Y_\lambda,Z_\lambda)
=
(\alpha_\lambda\beta_\lambda,
\alpha_\lambda\gamma_\lambda,
\beta_\lambda\gamma_\lambda)~,
\end{equation}
it belongs to the $N_\lambda=1$ curved region and therefore lies inside the tetrahedron-like 
(curved faces) structure.
The averaging over $\lambda$ resulting in the point $(X,Y,Z)$ corresponds to the
a convex combination of the $\lambda = 1, 2$ points and thus belongs to the regular tetrahedron.

Finally, it is important to note that no continuous distribution of hidden variables is
required to generate this region. Since the $\mathcal{SL}$ set is the convex hull of four
vertices in a three-dimensional space, any point $(X,Y,Z)$ in this tetrahedron
can be represented as a convex combination of at most four extremal points.
Equivalently, any $\mathcal{SL}$ model in this representation can be realised by a discrete
hidden-variable model with at most four values of $\lambda$. Fewer values suffice
for points lying on lower-dimensional subsets of the tetrahedron: one value for
the vertices, two for points on edges, and three for points on triangular faces.
Therefore, while Eq.~\eqref{eq:SL_average} allows for an arbitrary number of hidden-variable values,
the full $\mathcal{SL}$ region in the pyramid representation is already completely
generated in models with four $\lambda$'s.

\section{Number of Bell-like inequalities}
\label{sec:number}
\subsection{Barycentric coordinates}
From the previous section, it follows that
a point $(X_0, Y_0, Z_0)$ corresponds to an $\mathcal{SL}$ model if it belongs to the 
regular tetrahedron given by the vertices $V_1, V_2, V_3, V_4$.

Let us introduce normalised barycentric coordinates $\xi_i$ by
\begin{align}
X_0 &= \sum_{i=1}^4 \xi_i V^X_i~=\xi_1-\xi_2-\xi_3+\xi_4 ~,\nonumber\\
Y_0 &= \sum_{i=1}^4 \xi_i V^Y_i=\xi_1-\xi_2+\xi_3-\xi_4~, \nonumber\\
Z_0 &= \sum_{i=1}^4 \xi_i V^Z_i=\xi_1+\xi_2-\xi_3-\xi_4~,
\label{eq:baryc}
\end{align}
where superscripts denote the respective $X$, $Y$ and $Z$ coordinates of the vertices
and four barycentric coordinates are normalised as follows:
\begin{equation}
\xi_1+\xi_2+\xi_3+\xi_4=1~.
\label{eq:norm}
\end{equation}
The point $(X_0, Y_0, Z_0)$ is in the interior of the tetrahedron if 
\begin{equation}
\xi_i > 0~,
\qquad i=1,\dots,4~.
\label{eq:pos}
\end{equation}
It is easy to see that under Eq.~\eqref{eq:norm}, the four inequalities~\eqref{eq:pos} are equivalent to the following three chain inequalities:
\begin{align}
\xi_1\xi_2 & > 0~, \nonumber\\
\xi_2\xi_3 & > 0~, \nonumber\\
\xi_3\xi_4 & > 0~.
\label{eq:chain}
\end{align}
Noting the relation between the barycentric coordinates and 
$(X_0, Y_0, Z_0)$ obtained by inverting Eq.~\eqref{eq:baryc} and using the constraint of Eq. ~\eqref{eq:norm}, one gets:
\begin{align}
\xi_1 &= \tfrac14(1+X_0+Y_0+Z_0)~, \nonumber\\
\xi_2 &= \tfrac14(1-X_0-Y_0+Z_0)~, \nonumber\\
\xi_3 &= \tfrac14(1-X_0+Y_0-Z_0)~, \nonumber\\
\xi_4 &= \tfrac14(1+X_0-Y_0-Z_0)~.
\label{eq:xi}
\end{align}
Then, three inequalities~\eqref{eq:chain} in terms of $(X_0, Y_0, Z_0)$ follow:
\begin{align}
(1+Z_0)^2 & > (X_0+Y_0)^2~, \nonumber\\
(1-X_0)^2 & > (Y_0-Z_0)^2~, \nonumber\\
(1-Z_0)^2 & > (X_0-Y_0)^2~. 
\label{eq:SLquad3}
\end{align}
They are sufficient to test whether a point in the $(X, Y, Z)$ space is in the interior of the tetrahedron.
Note that the inequalities exclude the tetrahedron facets.

\vspace{0.2cm}
Finally, we address the question of whether three is the minimum number of independent inequalities needed to separate the $\mathcal{SL}$ region from the non-symmetric local ($\mathcal{\overline{SL}}$) region in the symmetry-reduced $(X,Y,Z)$ space.
Let $K\subset\mathbb{R}^3$ be a bounded convex set with non-empty interior.
If $K$ is written as the intersection of $k$ independent closed half-spaces, then each inequality
contributes a distinct bounding direction.
For $k\le 2$, the intersection is unbounded (a half-space, wedge, or slab), hence cannot equal a
bounded three-dimensional body.
Therefore, no bounded three-dimensional convex region can be defined by fewer than three independent
inequalities.
Since the $\mathcal{SL}$ region in the present reduced space is a tetrahedron, at least three inequalities (such as in Eq.~\eqref{eq:SLquad3}) are required to define its interior.

\vspace{0.2cm}
\subsection{Relation to the full $(3, 3, 2, 2)$ Bell polytope}

It is important to distinguish the reduced three-dimensional space used here from the full
$(3,3,2,2)$ Bell scenario in 36-dimensional conditional-probability space.
By explicit introduction of probability normalisation, the
36 conditional probabilities in the $(3,3,2,2)$ scenario
are reduced to $36-9=27$ independent parameters,
since each of the nine setting pairs carries one normalisation
constraint.

Imposing exchange symmetry given by Eq.~\eqref{eq:symm_g} identifies the six off-diagonal setting pairs into three symmetric pairs, while the three diagonal setting pairs $(0,0)$, $(1,1)$, and $(2,2)$ remain distinct. Thus, there are six independent combinations of settings in total.
Each of the three symmetric off-diagonal combinations contributes three independent probabilities after normalisation, for a total of $3\times3=9$ parameters.
Each of the three diagonal combinations contributes two independent probabilities after normalisation and the additional symmetry constraint $P(+1,-1 \mid q,q)=P(-1,+1 \mid q,q)$, for a total of $3\times2=6$ parameters.
Therefore, the dimension of the symmetry-reduced conditional-probability space is $9+6=15$.

One may further restrict the symmetric probability space by imposing the
no-signalling conditions discussed in Appendix~\ref{app:no-signalling}. For binary outcomes,
any conditional probability distribution can be parametrised by mean of the marginal distribution and correlators. In the exchange-symmetric case,
the six independent setting combinations are characterised by six
correlators,
\[
M_{00},\, M_{11},\, M_{22},\, M_{01},\, M_{02},\, M_{12},
\]
and by the nine independent means. The no-signalling
conditions require that the means at a given site
depend only on the local setting and not on the setting chosen at the distant site. 
Together with exchange symmetry, this reduces the nine
local-mean parameters to three. The six correlators remain unrestricted by
the no-signalling conditions apart from positivity bounds. Consequently,
the dimension of the exchange-symmetric no-signalling space is
\[
3 + 6 = 9~.
\]
Thus, imposing no-signalling reduces the dimension of the
exchange-symmetric conditional-probability space from 15 to 9.

Importantly, this 9-dimensional space still contains the full
$\mathcal{SL}$ polytope compatible with indistinguishability of sites.
The present work performs a further projection of this
9-dimensional probability space onto the
three-dimensional mixed-moment space $(X,Y,Z)$ of Eq.~\eqref{eq:mm}.
Under this projection, many distinct facets of the full polytope
collapse onto common bounding directions, and the $\mathcal{SL}$ region
assumes the simple geometric form of a regular tetrahedron.

In the full space $\mathcal{C}=\{P(a_1,a_2 \mid q_1,q_2)\}$ the symmetric local set $\mathcal{SL}$ is a high-dimensional convex
polytope with a rich facet structure.
Besides positivity constraints, there are two inequivalent non-trivial facet families, namely CHSH-type
inequalities and the $I_{3322}$ (Froissart) inequality~\cite{Froissart:1981ni, Brunner:2013est}.
When all relabelings are taken into account, the local polytope has 684 facet-defining inequalities
(36 positivity facets, 72 CHSH-type facets, and 576 $I_{3322}$-type facets).

Still in the 3D mixed-moment space, one can uniquely distinguish the $\mathcal{SL}$ models from the rest. 
The $\mathcal{SL}$ region is a regular tetrahedron; within this reduced affine space, only three inequalities are required to define its interior. Thus, the reduction $684\to 3$ reflects normalisation and symmetry reduction, and importantly, projection to a specially defined space of mixed moments.

The pyramid inequalities are thus significantly simpler for distinguishing the $\mathcal{SL}$ models when one restricts attention to models of indistinguishable sites. It does not, however, replace the full Bell-polytope description, which remains
necessary for arbitrarily different sites.

\section{Discussion}
\label{sec:physics}
The pyramid construction isolates a symmetry-reduced sector of the
bipartite Bell scenario with three measurement settings per site and
binary outcomes, i.e.\ the $(3,3,2,2)$ scenario.  For indistinguishable
sites, the exchange symmetry of the observed distribution reduces the six
off-diagonal correlators to the three coordinates
\[
X=M_{01}=M_{10}~,\qquad
Y=M_{02}=M_{20}~,\qquad
Z=M_{12}=M_{21}~.
\]
Three settings are the smallest number that produce a closed triangle of
pairwise off-diagonal correlations, $(0,1)$, $(0,2)$, and $(1,2)$.  This
makes it possible to test the mutual consistency, or ``correlation
transitivity'', of the three pairwise relations without using diagonal
correlators.  With only two settings, the corresponding symmetry reduction
leaves a single off-diagonal correlator and no analogous triangular
consistency condition (see Appendix~\ref{app:three_settings}).

\begin{figure}[!t]
\centering
\includegraphics[width=0.7\linewidth]{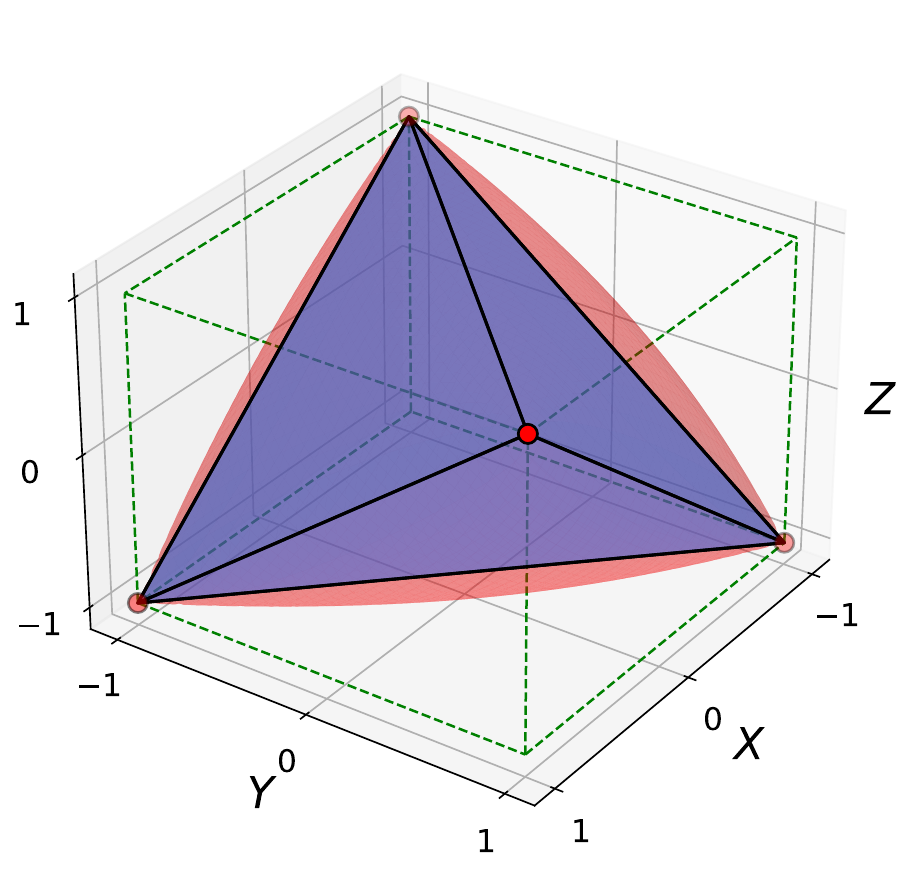}
\caption{
Geometric relation among the symmetric-local, symmetric-quantum, and
no-signalling sets in the symmetry-reduced mixed-moment space $(X,Y,Z)$.
The symmetric-local set $\mathcal{SL}$ is the regular tetrahedron (blue).
The symmetric-quantum set $\mathcal{SQ}$ is the larger elliptope (red),
whose curved boundary satisfies
$1+2XYZ-X^2-Y^2-Z^2=0$~.
The projected no-signalling set $\mathcal{NS}$ is the cube
$[-1,1]^3$ (green dashed edges).  Projection onto the three mixed moments
removes all cross-coordinate no-signalling constraints, leaving only the
individual correlator bounds.
}
\label{fig:SL+QNS}
\end{figure}

In this space, $\mathcal{SL}$ is the convex hull of the four deterministic
vertices
\[
(1,1,1)~,\quad (-1,-1,1)~,\quad (-1,1,-1)~,\quad (1,-1,-1)~.
\]
Membership in the closed tetrahedron, including its facets, is therefore
equivalent to the four barycentric conditions
\begin{align*}
1+X+Y+Z &\geq 0~, &
1-X-Y+Z &\geq 0~,\\
1-X+Y-Z &\geq 0~, &
1+X-Y-Z &\geq 0~.
\end{align*}
As shown in Sec.~\ref{sec:tetrahedron}, positivity of the barycentric
coordinates can also be combined into three strict quadratic chain
inequalities that test membership in the \emph{interior}.  The distinction
between the open interior and the closed tetrahedron is important when
points on the facets are included.

The interpretation of this test should also be kept explicit.  Equality of
the hidden-variable response laws at the two sites is stronger than exchange
symmetry of the observed probabilities alone.  A violation of a pyramid
inequality therefore rules out the symmetric-local class $\mathcal{SL}$ as
defined here, but does not, by itself, rule out every local hidden-variable
model with different response functions at the two sites.  Moreover, the
projection discards the marginals, the diagonal correlators, and the other
coordinates of the full behaviour.  The pyramid test is consequently a
complete test for compatibility with $\mathcal{SL}$ in the reduced space,
not a replacement for the full Bell-polytope analysis.

To compare $\mathcal{SL}$ with quantum correlations, an equally precise
definition of the reduced quantum class is required.  Tsirelson's theorem
for dichotomic observables gives a vector representation
\cite{Tsirelson:1980}
\[
M_{ij}=\mathbf{u}_{i}\mathbin{\cdot}\mathbf{v}_{j}~,
\]
where the vectors are real and of unit norm.  Observed exchange symmetry,
$M_{ij}=M_{ji}$, does not by itself imply that the two vector families can
be identified.  We therefore define a \emph{symmetric quantum realisation}
in the present construction by the additional common-vector condition
\[
\mathbf{u}_{i}=\mathbf{v}_{i}\qquad (i=0,1,2)~.
\]
It follows that
\[
M_{ij}=\mathbf{u}_{i}\mathbin{\cdot}\mathbf{u}_{j}=M_{ji}~,
\qquad M_{ii}=1~.
\]
Thus, equal setting labels give perfect correlations.  This is a defining
property of the class $\mathcal{SQ}$ used in the pyramid representation,
rather than a consequence of observed exchange symmetry alone.

The off-diagonal correlators are then
\[
X=\mathbf{u}_{0}\mathbin{\cdot}\mathbf{u}_{1}~,\qquad
Y=\mathbf{u}_{0}\mathbin{\cdot}\mathbf{u}_{2}~,\qquad
Z=\mathbf{u}_{1}\mathbin{\cdot}\mathbf{u}_{2}~,
\]
and hence the matrix
\[
G=
\begin{pmatrix}
1 & X & Y \\
X & 1 & Z \\
Y & Z & 1
\end{pmatrix}
\]
must be positive semidefinite.  Conversely, every positive-semidefinite
matrix of this form is the Gram matrix of three real unit vectors.  The
symmetric-quantum region is therefore
\[
\mathcal{SQ}
=
\left\{
(X,Y,Z)\in[-1,1]^3:
1+2XYZ-X^2-Y^2-Z^2\geq 0
\right\}~,
\]
the three-dimensional elliptope discussed in
Appendix~\ref{app:quantum_region}.

The inclusion $\mathcal{SL}\subseteq\mathcal{SQ}$ is transparent in this
representation: all four vertices of the symmetric-local tetrahedron
correspond to positive-semidefinite Gram matrices, and the elliptope is
convex.\footnote{At the level of the
projected coordinates $(X,Y,Z)$, every point of $\mathcal{SL}$ has an
alternative realisation as a convex mixture of the four deterministic
symmetric strategies.  This statement concerns the projected off-diagonal
correlators; it does not assert that an arbitrary stochastic realisation has
the same probabilities for the discarded diagonal setting pairs.}
The inclusion is strict.  For example,
\[
(X,Y,Z)=\left(\frac{1}{\sqrt{2}},\frac{1}{\sqrt{2}},0\right)
\]
lies on the elliptope boundary but violates
$1-X-Y+Z\geq0$~, and hence lies outside $\mathcal{SL}$.

The no-signalling projection is even simpler.  No-signalling constrains the
marginals but imposes no relation among the three retained correlators.
Indeed, for any symmetric assignment $M_{ij}\in[-1,1]$, the zero-marginal
behaviour
\[
P(a_1,a_2\mid i,j)
=\frac{1}{4}\left(1+a_1a_2M_{ij}\right)
\]
is positive, normalised, and no-signalling.  Consequently,
\[
\mathcal{NS}=[-1,1]^3~.
\]
This second inclusion is also strict: the no-signalling point
$(X,Y,Z)=(1,1,-1)$ has
$1+2XYZ-X^2-Y^2-Z^2=-4$ and therefore lies outside
$\mathcal{SQ}$.  Thus, for the projected classes defined above,
\[
\mathcal{SL}\subsetneq\mathcal{SQ}\subsetneq\mathcal{NS}~,
\]
as illustrated in Fig.~\ref{fig:SL+QNS}.  This separation is the
three-setting analogue of the familiar distinction between quantum and
post-quantum no-signalling correlations, of which PR-box correlations are
the standard example in the full behaviour space
\cite{Popescu:1994kjy,Brunner:2013est}.

Unlike in the popular the $(2,2,2,2)$ case,
the pyramid representation of the (3,3,2,2) case allows searches for breaking correlation transitivity~\ref{app:three_settings} and avoids using diagonal mixed moments, still preserving a high degree of simplicity.

The required symmetry structure suggests applications in particle and
nuclear physics whenever the two measurement wings admit a common
setting--outcome convention and approximately symmetric experimental
conditions.  Relevant contexts include spin-correlation measurements in
hyperon--antihyperon pairs~\cite{STAR:2025njp,Wu:2024mtj},
$\tau\bar{\tau}$ decays of the Higgs boson~\cite{Barr:2024djo},
$t\bar{t}$ production at the LHC~\cite{ATLAS:2023fsd}, Bell-type studies
of positronium decays~\cite{Moskal:2024fpv,Kumar:2023xpx}, and theoretical
proposals involving kaon and neutrino oscillations~\cite{Bramon:2004zp,Blasone:2015lya}.  In many of these settings, however,
the effective measurement choices are fixed or inferred from decay
kinematics rather than selected independently on each trial.  Such studies
can probe entanglement and quantum correlations, but they are not strict
Bell tests unless the assumptions of measurement independence, spatial
separation, and independently chosen settings are operationally satisfied.

\section{Summary}
\label{sec:summary}

This paper introduces the \emph{pyramid inequalities} for identifying
local models in the bipartite Bell scenario with
three measurement settings per site and binary outcomes,
i.e.\ the $(3,3,2,2)$ case.
Instead of analysing the full 36-dimensional conditional-probability
space---where the local polytope is known to possess 684 facet-defining
inequalities---we focus on a symmetry-reduced description appropriate
for indistinguishable sites.

In this formulation, the relevant observables are the mixed
off-diagonal correlators which define a three-dimensional mixed-moment space.
Within this space, the following results are obtained:

\begin{enumerate}[(i)]
\item \textbf{Single hidden-variable value case ($N_\lambda=1$).}
The accessible $\mathcal{SL}$ region forms a tetrahedron-like structure with
curved faces, generated by factorisable correlators
\[
X=\alpha\beta~,\quad Y=\alpha\gamma~,\quad Z=\beta\gamma~.
\]

\item \textbf{General $\mathcal{SL}$ models (convex mixtures).}
Averaging over the hidden variable renders the region convex.
The full $\mathcal{SL}$ set becomes the convex hull of four deterministic
extremal points,
\[
(1,1,1)~,\; (-1,-1,1)~,\; (-1,1,-1)~,\; (1,-1,-1)~,
\]
i.e.\ a regular tetrahedron with planar faces.

\item \textbf{Minimal inequality description.}
Although the tetrahedron has four facets, only three inequalities are required to describe its interior. Thus, within the symmetry-reduced Bell's $(3,3,2,2$ case, only three inequalities, the pyramid inequalities, are needed to fully define the interior of the $\mathcal{SL}$ body.

\item \textbf{Relation to the full $(3,3,2,2)$ polytope.}
The reduction from 684 facet inequalities in the full probability space to three inequalities in the mixed-moment space defining the interior of the $\mathcal{SL}$ body reflects
probability normalisation, symmetry reduction, and projection to a
specially constructed correlator subspace.
The pyramid inequalities thus provide a significantly simpler test
of locality in the symmetric setting, while not replacing the
full polytope description for arbitrary asymmetric distributions.

\item \textbf{Symmetric quantum and no-signalling structure.}
The symmetric quantum region in the $(X,Y,Z)$ space has a shape of
the elliptope with the boundary given by
\[
1 + 2XYZ - X^2 - Y^2 - Z^2 \ge 0~.
\]
Remarkably, only a single inequality defines the symmetric quantum region.
The non-signalling models cover the full $(X,Y,Z)$ space:
$\mathcal{NS} = [-1,1]^3~$~.

Thus, the hierarchy
\[
\mathcal{SL} \subsetneq \mathcal{SQ} \subsetneq  \mathcal{NS}~,
\]
appears in the $(X,Y,Z)$ space as a regular tetrahedron embedded in a strictly
larger curved convex body, itself contained within the no-signalling
cube.
\end{enumerate}

Unlike the standard one-dimensional representation of the $(2,2,2,2)$ case,
the pyramid representation of the (3,3,2,2) case allows searches for breaking correlation transitivity and avoids using diagonal mixed moments, still preserving a high degree of simplicity.

The pyramid representation assumes site indistinguishability, suggesting possible applications in particle and nuclear physics, where identical particles and symmetric experimental conditions arise naturally.

\section*{Acknowledgements}
The authors thank Jean-Daniel Bancal for critical comments.
This work was supported by the Polish National Science Centre grant 2018/30/A/ST2/00226 and by the Polish Minister of Science under
the ‘Regional Excellence Initiative’ program (project RID/SP/0015/2024/01),

\appendix

\section*{Appendices}

\section{Why Three Settings per Site? 
Gain over the $(2,2,2,2)$ Case}
\label{app:three_settings}

In the standard $(2,2,2,2)$ Bell scenario (two settings per site, binary outcomes), the symmetric local polytope is completely characterised 
by CHSH-type inequalities.
Quantum non-locality is typically quantified by the correlator $S$, which for quantum models is limited 
by the Tsirelson bound $2\sqrt{2}$.

In contrast, the $(3,3,2,2)$ scenario considered in this work
contains a new three-dimensional structure. This enriches physics potential and applications.

\textbf{Breaking Correlation Transitivity -- Triadic Frustration.}
The pyramid representation probes whether correlators for pairwise settings satisfy a
transitivity property.  To make this explicit, consider the three
measurement settings labelled \(0,1,2\).  In the symmetry-reduced
representation, the three mixed moments
\[
X=M_{01}~,\qquad
Y=M_{02}~,\qquad
Z=M_{12}~,
\]
yield relations between three distinct pairs of settings:
\[
0 \leftrightarrow 1~,\qquad
0 \leftrightarrow 2~,\qquad
1 \leftrightarrow 2~.
\]
They may be viewed as the three edges of a
triangle whose vertices correspond to the settings \(0,1,2\).

In the standard \((2,2,2,2)\) scenario, there are only two settings per site. After symmetry reduction, there is only one independent cross-correlation,
\[
X=M_{01}~,
\]
corresponding to a single edge connecting the two settings \(0\) and
\(1\). Since one has only a single link rather than a closed triangular
structure, no notion of correlation transitivity can arise.

In contrast, in the \((3,3,2,2)\) pyramid representation, the three
correlators \(X,Y,Z\) close into a loop. Symmetric local models
impose consistency conditions between these three edges. In
particular, if the correlations represented by two edges are strong,
then the third edge must also be sufficiently strong. 
The transitivity property is not imposed independently, but emerges from
Bell factorisation in symmetric local models\footnote{In the symmetry-reduced
representation, Bell's factorisation also leads to the
pyramid inequalities bounding the tetrahedral \(SL\) region.}
Symbolically, if
\[
0 \leftrightarrow 1
\quad\text{and}\quad
0 \leftrightarrow 2
\]
are both strongly correlated, then locality requires a non-zero
correlation also along
\[
1 \leftrightarrow 2~.
\]

Quantum models can violate this transitivity because quantum correlations
are not constrained by Bell's factorisation \eqref{eq:bell_factorisation}.
For example, for
\[
X=Y=\frac{1}{\sqrt2}\approx0.707~,
\]
the pyramid inequalities imply
\[
Z \ge 0.414
\]
for symmetric local models. However, quantum models can realise
\[
Z=0~.
\]
Geometrically, two sides of the triangle remain strongly correlated,
while the third side vanishes. The closed correlation loop is therefore
`frustrated', analogously to frustrated spin triangles in condensed
matter physics~\cite{MoessnerRamirez2006}.

Such frustration cannot occur in the \((2,2,2,2)\) case because there is
no closed correlation loop. Thus, the presence or absence of triadic
frustration provides a geometric and topological witness distinguishing
local from non-local behaviour. At least three
settings per site are required for such a closed-loop test, which is why
this phenomenon first appears in the \((3,3,2,2)\) scenario.

\textbf{Diagonal-Free Testing — No Identical Settings Required.}
In the $(2,2,2,2)$ scenario with indistinguishable sites, proving non-locality requires measuring diagonal moments $M_{00}$ and $M_{11}$ — cases where the exact same setting is chosen simultaneously at sites 1 and 2.

For indistinguishable sites, this creates potential experimental complications. In principle, identical particles measured with identical settings may exhibit bunching or anti-bunching (for bosons or fermions, respectively), which can distort the observed statistics. While such exchange effects are typically suppressed when the particles are well-separated and effectively distinguishable, they can re-emerge in realistic implementations where mode overlap, or detection ambiguity, is present. 
Moreover, requiring both sites to implement the same setting necessitates synchronised and effectively identical measurement apparatus, which can weaken the operational independence of the two sites and thereby blur the distinction between a genuinely bipartite measurement and a common measurement context applied to both subsystems.
Finally, diagonal measurements can enhance sensitivity to fair-sampling loopholes, since identical settings at both sites lead to correlated detection efficiencies, thereby amplifying selection biases in the subset of detected events.

The pyramid inequalities are built exclusively from off-diagonal moments $X,Y,Z$. The three settings 
$(0, 1, 2)$ per site ensure every cross-relation uses a different setting on each site ———diagonals are never needed to witness non-symmetric locality.

\section{Symmetric quantum region in the pyramid representation}
\label{app:quantum_region}

Here, we characterise the set of quantum models in the
symmetry-reduced mixed-moment space introduced in Sec.~\ref{sec:tetrahedron}.
The result follows from a general theorem due to Tsirelson~\cite{Tsirelson:1980}
on the structure of quantum correlators for dichotomic observables; see also the review~\cite{Brunner:2013est}.

\textbf{Tsirelson representation of quantum correlators.}
Consider a bipartite Bell experiment as defined in Sec.~\ref{sec:setup}, with
measurement settings $q_1,q_2\in\{0,1,2\}$ and binary outcomes
$a_1,a_2\in\{-1,+1\}$. The correlators are defined as
\begin{equation}
M_{q_1q_2} = \langle a_1 a_2 \rangle_{q_1\,q_2}~.
\end{equation}
A fundamental result due to Tsirelson~\cite{Tsirelson:1980} states that for any quantum
realisation of such correlations, there exist unit vectors
$\mathbf{u}_{q_1}$ and $\mathbf{v}_{q_2}$ in a real Hilbert space such that
\begin{equation}
\label{eq:unit_vec}
M_{q_1q_2} = \mathbf{u}_{q_1} \cdot \mathbf{v}_{q_2}~.
\end{equation}
This representation is general and does not depend on the number of measurement settings.

\textbf{Reduction to the symmetry-reduced coordinates.}
In the symmetry-reduced description introduced in Sec.~\ref{sec:tetrahedron}, we consider the three mixed moments
\begin{equation}
X=M_{01}~,\qquad
Y=M_{02}~,\qquad
Z=M_{12}~.
\end{equation}

Because sites 1 and 2 are assumed to be indistinguishable, the off-diagonal correlators satisfy
\begin{equation}
M_{q_1 q_2} = M_{q_2 q_1}~.
\end{equation}
This observed exchange symmetry means that the bipartite cross-correlation matrix is symmetric, but by itself it does not yet reduce the quantum description to a single family of vectors. In a general quantum realisation, Eq.~\eqref{eq:unit_vec}, has 
one family of unit vectors associated with site 1 and another associated with site 2. To obtain the reduced three-dimensional description used here, we therefore make the additional assumption that the same setting label corresponds to the same effective correlation direction on both wings. In this respect, the assumption plays, on the quantum side, a role analogous to the hidden symmetry Eq.~\eqref{eq:symm_b} in the symmetric local construction: there, Eq.~\eqref{eq:symm_b} replaces two symmetric local response families by a single setting-indexed response law, while here the quantum assumption replaces the two vector families by one common setting-indexed family. Under this common-vector condition, the three correlators \((X,Y,Z)\) can be written as
\[
X = \mathbf{u}_0 \cdot \mathbf{u}_1~, \quad Y = \mathbf{u}_0 \cdot \mathbf{u}_2~, \quad 
Z = \mathbf{u}_1 \cdot \mathbf{u}_2~.
\]
Accordingly, the symmetric quantum region is characterised by the Gram matrix of the three vectors \(u_0,u_1,u_2\), and hence by the elliptope derived below.

\textbf{Gram matrix characterisation.}
Define the Gram matrix
\begin{equation}
G=
\begin{pmatrix}
1 & X & Y \\
X & 1 & Z \\
Y & Z & 1
\end{pmatrix}.
\end{equation}
This matrix contains all scalar products of the three unit vectors
$u_0,u_1,u_2$.
A necessary and sufficient condition for such vectors to exist
is that the Gram matrix is positive semidefinite,
\begin{equation}
G\succeq0~.
\end{equation}
For a symmetric $3\times3$ matrix, this is equivalent to the
non-negativity of all principal minors.

\textbf{Resulting inequalities.}
The $2\times2$ principal minors yield
\begin{equation}
1-X^2\ge0~,\qquad
1-Y^2\ge0~,\qquad
1-Z^2\ge0~,
\end{equation}
which implies
\begin{equation}
|X|\le1~,\qquad |Y|\le1~,\qquad |Z|\le1~.
\end{equation}

The determinant condition gives
\begin{equation}
\det G
=1+2XYZ-X^2-Y^2-Z^2 \ge0~.
\end{equation}

Therefore, the symmetric quantum region in the pyramid representation is
\begin{equation}
Q=\left\{(X,Y,Z)\in[-1,1]^3 \; \middle| \;
1+2XYZ-X^2-Y^2-Z^2\ge0 \right\}~.
\end{equation}

Geometrically, this set is the three-dimensional \emph{elliptope},
a convex curved body bounded by the elliptope surface
\begin{equation}
1+2XYZ-X^2-Y^2-Z^2=0~.
\label{Eq:elliptope}
\end{equation}

In this respect, it is instructive to present a quantum-mechanical example in which $X, Y, Z$ points lie on the elliptope surface. Let us consider two photons emitted back-to-back along the $z$-axis, the first photon propagates to the left ($z<0$), the second to the right ($z>0$), with wave-function:
\begin{equation}
\left\vert \Psi\right\rangle = \frac{1}{\sqrt{2}}\left( \left\vert HH\right\rangle + \left\vert VV\right\rangle \right)~,
\end{equation}
where $H$ and $V$ denote horizontal (along the $x$-axis) and vertical (along the $y$-axis) polarizations, respectively.

Two polarisers are placed, one on each site, with three possible setting angles $\theta_{0,1,2}$, corresponding to the three setting labels $q_{0,1,2}$. When the same setting is chosen on both sites, identical outcomes are obtained (both photons are either transmitted or rejected).

For different angles, the correlations are given by
\begin{equation}
X = \cos\bigl(2(\theta_0 - \theta_1)\bigr)~, \quad
Y = \cos\bigl(2(\theta_0 - \theta_2)\bigr)~, \quad
Z = \cos\bigl(2(\theta_1 - \theta_2)\bigr)~.
\label{eq:qmm}
\end{equation}
It is easy to see that $(X,Y,Z)$ points given by Eq.~(\eqref{eq:qmm})
satisfy Eq.~(\eqref{Eq:elliptope}) for any choice of $\theta_{0,1,2}$.

\section{No-signalling region in the pyramid representation}
\label{app:no-signalling}

Here, we characterise the set of non-signalling models in the
symmetry-reduced mixed-moment space introduced in Sec.~\ref{sec:tetrahedron}.

\textbf{Definition of no-signalling models.}
A probabilistic model specified by the conditional probability distribution
$P(a_1,a_2 \mid q_1,q_2)$ is said to satisfy the no-signalling condition if the marginal distribution at each site does not depend on the measurement setting chosen at the other site. This is expressed by the
conditions
\begin{equation}
\sum_{a_2} P(a_1,a_2 \mid q_1,q_2)
=
\sum_{a_2} P(a_1,a_2 \mid q_1,q_2')
\quad \text{for all } q_2~\text{and}~q_2'~,
\label{eq:m1}
\end{equation}
\begin{equation}
\sum_{a_1} P(a_1,a_2 \mid q_1,q_2)
=
\sum_{a_1} P(a_1,a_2 \mid q_1',q_2)
\quad \text{for all } q_1~\text{and}~q_1'~.
\label{eq:m2}
\end{equation}

Importantly, the no-signalling principle alone does not uniquely
characterise quantum correlations.
Popescu and Rohrlich showed that there exist hypothetical correlations
that respect the no-signalling condition but are beyond quantum correlations~\cite{Popescu:1994kjy}.
The best-known example is the PR box, which violates the CHSH
inequality up to its algebraic maximum while still satisfying
the no-signalling constraints (see also the review~\cite{Brunner:2013est}).
Consequently, the hierarchy of physically relevant model sets is
\begin{equation}
\mathcal{SL} \subsetneq \mathcal{SQ} \subsetneq \mathcal{NS}~.
\end{equation}

\textbf{Probabilistic framework of this paper.}
At location $1$, the setting is
$q_1\in\{0,1,2\}$ and the outcome is
$a_1\in\{-1,+1\}$.
At location $2,$ the setting is
$q_2\in\{0,1,2\}$ and the outcome is
$a_2\in\{-1,+1\}$.

A behaviour is specified by conditional probabilities
\begin{equation}
P(a_1,a_2 \mid q_1,q_2)~.
\end{equation}

For binary outcomes, any probability distribution can be written in the form
\begin{equation}
P(a_1,a_2 \mid q_1,q_2)
=
\frac{1}{4}
\left(
1 + a_1 A^{(1)}_{q_1 q_2}
+ a_2 A^{(2)}_{q_1 q_2}
+ a_1 a_2 M_{q_1 q_2}
\right),
\end{equation}
where
\begin{align}
A^{(1)}_{q_1 q_2} &= \langle a_1 \rangle_{q_1q_2}~, \\
A^{(2)}_{q_1 q_2} &= \langle a_2 \rangle_{q_1q_2}~, \\
M_{q_1 q_2} &= \langle a_1 a_2 \rangle_{q_1q_2}~.
\end{align}

\textbf{No-signalling constraints.}
No-signalling requires that the marginal distribution at one location
does not depend on the setting at the other location, Eqs.~\eqref{eq:m1} and~\eqref{eq:m2}.
Using the correlator expansion, one finds
\begin{equation}
\sum_{a_2} P(a_1,a_2 \mid q_1,q_2)
=
\frac{1}{2}
\left(
1 + a_1 A^{(1)}_{q_1 q_2}
\right),
\end{equation}
which is independent of $q_2$ only if
\begin{equation}
A^{(1)}_{q_1 q_2} = A^{(1)}_{q_1}~.
\end{equation}
Similarly,
\begin{equation}
\sum_{a_1} P(a_1,a_2 \mid q_1,q_2)
=
\frac{1}{2}
\left(
1 + a_2 A^{(2)}_{q_1 q_2}
\right),
\end{equation}
which is independent of $q_1$ only if
\begin{equation}
A^{(2)}_{q_1 q_2} = A^{(2)}_{q_2}~.
\end{equation}

Thus, the no-signalling principle restricts only the symmetric local
expectation values, which may depend on the symmetric local settings
$q_1$ and $q_2$ but not on the distant ones.
Importantly, the mixed moments $M_{q_1 q_2}$ do not enter the marginal
distributions and are therefore not constrained by the
no-signalling conditions.

Positivity of probabilities implies
\begin{equation}
|M_{q_1q_2}| \le 1
\qquad
\text{for all } q_1,q_2\in\{0,1,2\}~.
\end{equation}

\textbf{Projection to $(X,Y,Z)$.}
The full no-signalling polytope lives in the space of all
mixed moments $M_{q_1q_2}$ together with the symmetric local expectations
$A_{q_1}$ and $B_{q_2}$.
However, in the projection onto the three variables
\[
(X,Y,Z)=(M_{01},M_{02},M_{12})~,
\]
the remaining correlators and symmetric local expectations do not appear.
They may therefore be chosen freely provided that the positivity
conditions are satisfied.

\textbf{Realisation of the cube.}
It remains to show that every point of the cube can be realised by
a no-signalling behaviour.
For any chosen values of $(X,Y,Z)$ in $[-1,1]^3$ consider the
probability distribution
\begin{equation}
P(a_1,a_2 \mid q_1,q_2)
=
\frac{1}{4}
\left(
1 + a_1 a_2 M_{q_1 q_2}
\right),
\end{equation}
with
\[
M_{01}=X~, \qquad
M_{02}=Y~, \qquad
M_{12}=Z~,
\]
and all other $M_{q_1q_2}$ chosen arbitrarily within $[-1,1]$.
This distribution has vanishing symmetric local expectations and therefore
satisfies the no-signalling constraints.
For $|M_{q_1q_2}|\le1$ all probabilities are non-negative and
properly normalised.
Thus, every point of the cube corresponds to a valid
no-signalling model.

\textbf{No-signalling region.}
The no-signalling region in the pyramid representation is therefore
\begin{equation}
-1 \le X \le 1~, \qquad
-1 \le Y \le 1~, \qquad
-1 \le Z \le 1~.
\end{equation}

Geometrically,
\begin{equation}
\mathcal{NS} = [-1,1]^3~,
\end{equation}
i.e.\ the full cube.

\bibliographystyle{elsarticle-num}
\bibliography{references}

\end{document}